\documentclass[%
 reprint,
 amsmath,amssymb,
 aps,
]{revtex4-1}
\usepackage{float}
\usepackage{graphicx}
\usepackage{dcolumn}
\usepackage{bm}
\usepackage{color}
\usepackage{braket}
\usepackage{amsmath,amssymb}
\usepackage{subfig}
\begin{document}

\preprint{APS/123-QED}

\title{A Supersymmetric Approach to the Problem of Micro-bending Attenuation in Optical Waveguides}

\author{Stuart Ward$^{1,2}$}
\author{Rouzbeh Allahverdi$^1$}%
\author{Arash Mafi$^{1,2}$}
\email{mafi@unm.edu}
\affiliation{$^1$Department of Physics \& Astronomy, University of New Mexico, Albuquerque, NM 87106, USA.\\
$^2$Center for High Technology Materials, University of New Mexico, Albuquerque, NM 87106, USA.
}

\date{\today}

\begin{abstract}
Micro-bending is a well-known source of loss in optical waveguides. By treating the micro-bending as a stochastic process, the problem of loss mitigation can be modeled in terms of a Fokker-Planck equation. Given an initial refractive index profile, and taking micro-bending into account, we develop a formalism to derive a new refractive index profile which potentially results in less loss. Our formalism is based on applying the techniques of Supersymmetric Quantum Mechanics to a Fokker-Planck equation that is associated with a particular refractive index profile. We derive a non-linear differential equation, whose solutions determine whether an index profile can undergo a supersymmetric transformation that results in less loss. As an explicit example, we consider a monomial index profile. We show that there exists a range of values for the monomial exponent which results in the new index profile having less loss.
\end{abstract}

\maketitle

\section{Introduction}
The field of Supersymmetric Quantum Mechanics (SUSY-QM) arose from the study of dynamical symmetry breaking in supersymmetric field theories~\cite{Witten1}. SUSY-QM has become a field of study in it's own right~\cite{Witten2,Cooper1,Cooper2,Cooper3,Suk85b,Gango}, due to the many techniques that it has provided to quantum mechanics. Such techniques allow for the classification of Hamiltonians possessing identical energy spectra~\cite{Jafarizadeh,Dutt}, an understanding of why the WKB approximation~\cite{Dutt2,Dutt3} is exact for some potentials, and the classification of families of Hamiltonians which have identical scattering and reflection coefficients~\cite{scatter,Sukumar2}. Beyond the applications provided to quantum mechanics, SUSY-QM has been utilized in the design of optical systems \cite{Chumakov,Ganainy2012,Miri2013,Miri2013-2,Heinrich2014,Heinrich2014-2,Heinrich2014-3,Miri2014,Ganainy2015,Walasik2018,Midya2018,Midya2019,Walasik2019,Zhong2019,Hokmabadi2019}, and in the study of statistical mechanics~\cite{Junker,Morales,Polotto,PaSo82,Zirnbauer,Efetov,PaSo79} through the correspondence between the imaginary-time Schr\"{o}dinger equation and the Fokker-Planck equation (FPE).

The techniques of SUSY-QM are based on relating two partner Hamiltonians, $H_{-}$ and $H_{+}$, through a function $W(x),$ which is referred to as the superpotential. The asymptotic behavior of the superpotential dictates the relationship between the spectra of the two partner Hamiltonians. The partner Hamiltonians either have identical energy spectra ($\text{Spec}\{H_{-}\} = \text{Spec}\{H_{+}\}$), or the spectra will be identical with the exception of a missing zero energy groundstate ($\text{Spec}\{H_{-}\} = \text{Spec}\{H_{+}\}+\{0\}$). The process of deriving one partner Hamiltonian from the other, through the corresponding superpotential, is referred to as performing a SUSY transformation.

Micro-bending in an optical waveguide occurs when unwanted random bends are introduced to the waveguide during its production and processing~\cite{Gardner,Jay}. Such bends are small, but introduce loss in the form of leakage of the light rays. This is due to the contact angle of the rays no longer being greater than the critical angle required for total internal reflection. In this paper, we expand upon our previous work~\cite{Paper} of employing the machinery of SUSY-QM to address the problem of mitigating loss induced by micro-bending. We take the approach of casting the problem in the form of a stochastic model by treating the bending as a random variable~\cite{Rousseau,Arnaud,ArnaudII,ArnaudIII,Han}. We derive an associated FPE for a given refractive index profile, and then utilize the correspondence of the imaginary-time Schr\"{o}dinger equation with the FPE. This allows us to perform a SUSY transformation on the original FPE to derive a new FPE which has an additional ground state eigenvalue, and hence a mode with a lower decay constant. This FPE is then related to its corresponding refractive index profile, which will result in less loss compared to the original index profile. We derive a necessary condition for the existence of a SUSY transformation which leads to lower loss. As an explicit example, we consider the monomial refractive index profile and show how the value of the exponent determines whether an index profile resulting in less loss can be derived.

The rest of this paper is organized as follows. In Section II, we go through background material on SUSY-QM and its applications to optics and stochastic processes. In Section III, we discuss the micro-bending problem, derive a FPE that describes it, and cast it in the form of a Schr\"{o}dinger-like equation. In Section IV, we apply this formalism to a monomial refractive index profile and derive the necessary condition required for the derived index to display lower loss. We conclude the paper in Section V. We discuss some details related to the derivation of FPE in the Appendix. There we also provide a short example of using the method of isospectral deformation which, based on our developed formalism, allows for the classification of families of index profiles that exhibit equal loss.

\section{Background}

\subsection{Supersymmetric Quantum Mechanics} 
Consider the following pair of Hamiltonians, written in terms of general creation and annihilation operators,

\begin{equation}
\label{eq:Hamiltonian}
H_{-} \equiv A^{\dagger}A ~ ~ ~ , ~ ~ ~ H_{+} \equiv AA^{\dagger}.
\end{equation}
The creation and annihilation operators are defined as
\begin{equation}
\label{eq:A}
A =\frac{\hbar}{\sqrt{2m}}\frac{d}{dx} + W(x) ~ ~ ~ , ~ ~ ~ A^{\dagger} = -\frac{\hbar}{\sqrt{2m}}\frac{d}{dx} + W(x),
\end{equation}
where $W(x)$ is a real valued function known as the Superpotential. Plugging Eq.~(\ref{eq:A}) into Eq.~(\ref{eq:Hamiltonian}) gives 
\begin{equation}
H_{\pm} = -\frac{\hbar^2}{2m}\frac{d^2}{dx^2} + V_{\pm}(x),
\end{equation}
where the potentials $ V_{+}(x)$ and $ V_{-}(x)$ are known as the SUSY partner potentials. The partner potentials are given in terms of the superpotential as
\begin{equation}
\label{eq:vpm}
V_{\pm}(x) = W(x)^2 \pm \frac{\hbar}{\sqrt{2m}}\frac{dW(x)}{dx}.
\end{equation}

Depending on the asymptotic behavior of the superpotential $W(x)$, a normalizable ground state of the Hamiltonian $H_{-}$, with energy eigenvalue equal to zero ($H_{-}\psi_{-}^0 = 0$), can be obtained by setting $A\psi_{-}^0 = 0.$ This yields, up to a normalization constant,
\begin{equation}
\psi_{-}^0 \sim \exp\big(-\frac{\sqrt{2m}}{\hbar}\int W(x)dx\big).
\end{equation}
For $\psi_{-}^0$ to be normalizable, the superpotential must take on a positive value at $+\infty$ and a negative value at $-\infty$. When this is the case we say that SUSY is unbroken. Note that in the case of unbroken-SUSY, a superpotential can be derived from the normalized ground state wavefunction via 
\begin{equation}
\label{eq:susyground}
W = -\frac{\hbar}{\sqrt{2m}}\frac{\psi_{0}'}{\psi_{0}}. 
\end{equation}
If $W(x)$ does not allow for a normalizable ground state, the superpotential is said to exhibit broken SUSY. When this is the case, we will see that this results in the spectra being identical for $H_-$ and $H_+$.

We can now relate the spectra of the Hamiltonians as follows. Let $H_{+}$ act on $A\psi_{-}^n$, where $\psi_{-}^n$ is the nth eigenstate of the Hamiltonian $H_{-}$. We find an isospectral relationship between the two Hamiltonians, namely
\begin{eqnarray}
H_{+}(A\psi_{-}^n) = AA^{\dagger}A\psi_{-}^n = A(H_{-}\psi_{-}^n) = E_{-}^nA\psi_{-}^n \, . \nonumber
\end{eqnarray}
Similarly, for $H_{-}$ acting on $A^{\dagger}\psi_{+}^n$, we have
\begin{eqnarray}
H_{-}(A^{\dagger}\psi_{+}^n) = A^{\dagger}AA^{\dagger}\psi_{+}^n = A(H_{+}\psi_{+}^n) = E_{+}^n A^{\dagger}\psi_{+}^n \, . \nonumber
\end{eqnarray}
This gives an exact isospectral relation between the two partner Hamiltonians in the case of broken SUSY. For the case of unbroken SUSY, we see that the ground state $\psi_{-}^0$ cannot be mapped to a state of the $H_{+}$ Hamiltonian. Thus there exists a one-to-one mapping between the excited eigenstates of $H_{-}$ and $H_{+}$, but the $H_{-}$ Hamiltonian now contains an extra state that does not get mapped to $H_{+}$. Since this extra state is the ground state, it has the lowest energy eigenvalue in the spectra.

\subsection{Supersymmetry in Guided Wave Optics}

SUSY-QM has been used for many applications in the study of guided wave optics~\cite{ChrisI,ChrisII,ChrisIII,HeinrichI,Sergey,Stefano,Laba,Walasik}. Here, we provide a brief overview of the connection between SUSY-QM and wave optics. Although we will be using a ray picture of light propagation to analyze the problem of micro-bending attenuation, we include this section in order to make the use of SUSY-QM to the study of waveguides more apparent.

There is a close analogy between the Schr\"{o}dinger equation for a particle in a potential $V(x)$ and light propagation in an optical waveguide. To establish the analogy, consider the Helmholtz equation for electromagnetic wave propagation along the z-direction, in a z-invariant (and y-invariant) dielectric waveguide,
\begin{equation}
\label{eq:helmholtz}
\dfrac{\partial^2A(x)}{\partial x^2}+n^2(x)k_0^2A(x)=\beta^2A(x).
\end{equation}
Here, $A(x)$ is the transverse profile of the (TE for simplicity) electric field $E(x,y,z,t)=A(x)\exp(i \beta z-i\omega_0 t)$, which is assumed to propagate freely in the $z$ direction, $\beta$ is the propagation constant,
$n(x)$ is the refractive index of the waveguide, $x$ is the confining transverse dimension, $\omega_0=ck_0$, and 
$k_0=2\pi/\lambda$, where $c$ is the speed of light in vacuum. Equation~(\ref{eq:helmholtz}) is an eigenvalue problem in 
$\beta^2$ and guided modes are those solutions (eigenfunctions) with $\beta^2>n^2_{\rm cl}k_0^2$.

The one-to-one analogy between the Schr\"{o}dinger equation
\begin{equation}
\label{eq:schrodinger}
-\dfrac{\hbar^2}{2m}\dfrac{\partial^2A(x)}{\partial x^2}+V(x)A(x)=EA(x),
\end{equation}
and the Helmholtz equation Eq.~(\ref{eq:helmholtz}) for optical wave propagation along the z-direction, in a 
longitudinally (z-direction) invariant and transversely (x-direction) varying waveguide, can be 
established by making the following identifications:
\begin{eqnarray}
V(x)&=&\dfrac{\hbar^2k_0^2}{2m}\Big(n_0^2-n^2(x)\Big) \, , \\
E&=&\dfrac{\hbar^2k_0^2}{2m}\Big(n_0^2-n^2_{\rm eff}\Big) \, ,
\label{eq:schrodinger2}
\end{eqnarray}
where $n_{\rm eff}=\beta/k_0$, and $n_0$ is any constant refractive index that can be conveniently chosen to set $E=0$ for unbroken SUSY. 
Guided waves with $n_{\rm eff}>n_{\rm cl}$ are equivalent to electronic bound-states in Eq.~(\ref{eq:schrodinger}).

For simplicity in establishing the analogy in optics, it is convenient to set $\hbar/\sqrt{2m}=L_c$, where $L_c$ is an arbitrary length scale 
chosen for each particular problem for convenience (e.g. waveguide width). This results in the following form for the waveguide structure
\begin{eqnarray}
\label{eq:poten}
&& V_\pm(x)=k_0^2L^2_c\Big(n_0^2-n_\pm^2(x)\Big)=W^2(x)\pm L_c \frac{dW(x)}{dx} \, ,\\
&& E_\pm=k_0^2L^2_c\Big(n_0^2-n^2_{\rm eff}\Big) \, . 
\end{eqnarray}
%

In a typical waveguide structure, $n_0$ is chosen to be the effective refractive index of the principle mode with the largest effective index
related to $n_-(x)$ index profile, such that the ground state energy vanishes for $V_-(x)$. We can therefore write Eq.~(\ref{eq:poten}) as
\begin{eqnarray}
&& n_\pm^2(x)=n_0^2+(k_0L_c)^{-2}\Big(-W^2(x)\mp L_c \frac{dW(x)}{dx}\Big) \, ,\\
&& \beta^2_\pm=\beta^2_0-E_\pm/L^2_c \, . 
\end{eqnarray}

The above correspondence shows how the techniques of SUSY-QM can be imported to the study of guided wave optics.

\subsection{Supersymmetry and Stochastic Processes}

The relationship between SUSY-QM and classical stochastic dynamics has been well established~\cite{Rosu,PolottoII,Bern}. For our purposes we review the connection between the Fokker-Planck equation and SUSY-QM.

Consider the set of Langevin equations of the following form
\begin{equation}
\label{eq:LDER}
\frac{\partial x_i}{\partial t} = A_i(\textbf{x},t) + \sum_{k}B_{ik}(\textbf{x},t)\zeta_k(t),
\end{equation}
with  $\{x_{i}\} = \textbf{x}.$ Taking the mean and auto-correlation to be defined as
\begin{eqnarray}
\big<\zeta_k(t)\big> &= &0 \, , \nonumber \\
\big <\zeta_{i}(t_{1})\zeta_{j}(t_{2}) \big > & = & 2D\delta_{ij}\delta{(t_{1}-t_{2})}
\, ,
\end{eqnarray}
the corresponding Fokker-Planck equation~\cite{FP} is given by
\begin{eqnarray}
\label{eq:FPDER}
&& \frac{\partial P(\textbf{x},t)}{\partial t} =  \, \nonumber \\ 
&& - \sum_i\frac{\partial}{\partial x_i}\{[A_i(\textbf{x},t)+ D\sum_{jk}B_{jk}(\textbf{x},t)\frac{\partial B_{ik}(\textbf{x},t)}{\partial x_i}]P\} \, \nonumber \\
&& + D\sum_{ij}\frac{\partial^2}{\partial x_i \partial x_j}\{[\sum_kB_{ik}(\textbf{x},t)B_{jk}(\textbf{x},t)]P\} \, .
\end{eqnarray}

For the case of one spatial and one time dimension the FPE can be written as
\begin{eqnarray}
\dfrac{\partial}{\partial t}P(x,x_0;t) = \dfrac{D}{2}\dfrac{\partial^2}{\partial x^2}P(x,x_0;t)-\dfrac{\partial}{\partial x}F(x)P(x,x_0;t) \, . \nonumber \\
\, 
\label{eq:FokkerPlanck1}
\end{eqnarray}
This is a diffusion equation, where $D/2$ is the diffusion constant and $F$ is the drift coefficient. This equation can be transformed to a Schr\"{o}dinger-like equation if we let
\begin{equation}
P(x,x_0;t)=\exp\Big\{-\dfrac{1}{D}[U(x)-U(x_0)]\Big\}~K(x,t), 
\end{equation}
where
\begin{equation}
U(x)=-\int_0^x~dz~F(z).
\end{equation}
The transformed equation is given by
\begin{equation}
\label{eq:drift1}
-D\dfrac{\partial}{\partial t}K(x,t) = 
\Big(\dfrac{-D^2}{2}\dfrac{\partial^2}{\partial x^2} + W^2(x) + \dfrac{D}{\sqrt{2}} W^\prime(x)\Big)K(x,t),
\end{equation}
where $F(x)=\sqrt{2}W(x)$. Equation~(\ref{eq:drift1}) can be identified with the Schr\"{o}dinger equation if $D=\hbar/\sqrt{m}$ and we take the time to be imaginary. Clearly, the SUSY partner to this equation can be written by exchanging $W^\prime$
with $-W^\prime$ in Eq.~(\ref{eq:drift1}).
 
The eigenstates of the SUSY partner Hamiltonians for Eq.~(\ref{eq:drift1}) satisfy the following equation
\begin{equation}
H_\pm \phi^\pm_n(x) = \lambda^\pm_n \phi^\pm_n(x),
\end{equation}
where
\begin{equation}
H_\pm = \dfrac{-D}{2}\dfrac{\partial^2}{\partial x^2} + W^2(x) \pm \dfrac{D}{\sqrt{2}} W^\prime(x).
\end{equation}
Similar to the Schr\"{o}dinger equation, where the eigenstates of the Hamiltonian evolve with a phase proportional to their energy,
the eigenstates of the Hamiltonian in Eq.~(\ref{eq:drift1}) decay with a decay constant proportional to the eigenvalues $\lambda^\pm$.
As such, we can write a spectral representation for $P(x,x_0;t)$ as
\begin{align}
&P(x,x_0;t)  =  \exp\Big\{-\dfrac{1}{D}[U_\pm(x)-U_\pm(x_0)]\Big\} \, \nonumber
\\
&\times \sum^\infty_{n=0} \exp\Big\{-\dfrac{1}{D} t \lambda^\pm_n\Big\}
\phi^\pm_n(x)\phi^\pm_n(x_0) \, .
\end{align}
The presence of SUSY implies that all decay constants are positive. For broken SUSY, all $\lambda^+_n$ are equal to their partners
$\lambda^-_n$, while for unbroken SUSY the same is true except $\lambda^-_0$ has no counterpart in $H_+$. So we see, given that the system exhibits unbroken SUSY, there exists a distribution associated with $\lambda^-_0$ which does not decay over time. This will be the key feature we will utilize in deriving an alternative refractive index which mitigates micro-bending loss of the lowest mode (ground state) in the waveguide.

\section{Optical waveguide with micro-bending} 
In this section we show how, in the presence of microbending, a refractive index profile can be related to a Fokker-Planck equation. From here the Fokker-Planck equation is brought into the same form as an imaginary time Schr\"{o}dinger equation. This allows for a SUSY transformation to be performed, resulting in a new Fokker-Planck equation. We then show how to reverse the process and derive the refractive index profile from a corresponding Fokker-Planck equation.

\subsection{Derivation of the Fokker-Planck Equation}

For our model, we assume a ray picture for light propagation through a slab waveguide, which is confined in the $y$ direction and longitudinal in the $z$ direction. The results presented here can be readily generalized to model a fiber in two dimensions. 

The micro-bending effect can be modeled by taking the optical waveguide to be bent in the $y$-$z$ plane with a local bending radius $R(z)$. The local bending radius $R(z)$ is the inverse of the local curvature given by 
\begin{equation}
\kappa(z) = R^{-1}(z) = \Big|\frac{d^2y}{dz^2}\Big|\Big[ 1+(\dfrac{dy}{dz})^{2}\Big]^{-3/2}.
\end{equation}

\begin{figure}[h]
\includegraphics[width=0.35\textwidth]{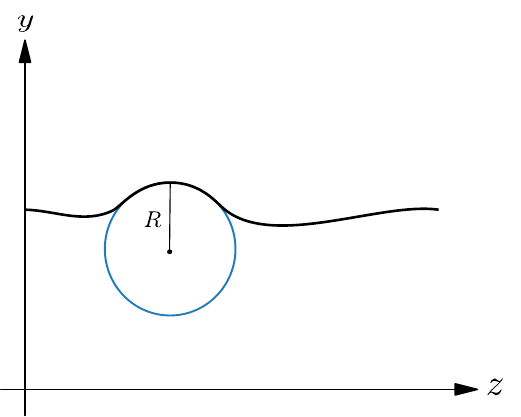}
\caption{Local bending radius of a curve.}
\label{fig:curvature}
\end{figure}

The bending radius along a curve is illustrated in Fig.~\ref{fig:curvature}.

In the absence of micro-bending, the refractive index profile for a graded index waveguide takes the general form

\begin{equation}
\label{eq:indexbending1}
n = n_{0}\big(1-U(y)\big).
\end{equation}

A commonly used graded index profile is the quadratic profile, which is defined by choosing $U(y) =\Delta (y/y_{c})^2.$ We will make use of this choice of profile in later sections, but for now we keep $U(y)$ as a general function.

It has been shown that in the study of curved waveguides, a conformal transformation may be applied to the waveguide which results in the addition of a linear term to the refractive index~\cite{MafiAcoustic,Heiblum,Menachem,Sheehan}. This is illustrated for a waveguide with constant index of refraction in Fig~2.

\begin{figure}[h]
\includegraphics[width=0.23\textwidth]{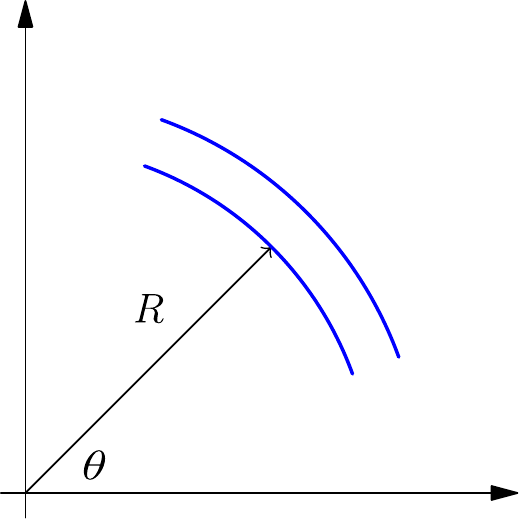}
\includegraphics[width=0.23\textwidth]{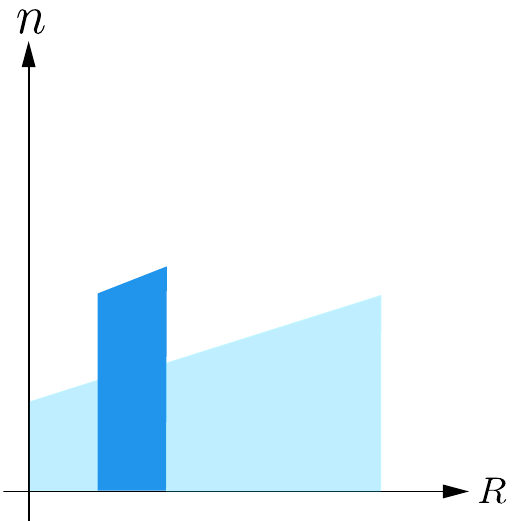}
\caption{A bent waveguide and its corresponding titled refractive index.}
\label{fig:conformal}
\end{figure}

Thus, in the presence of micro-bending, we can treat the bent waveguide as a straight waveguide with a modified index of refraction, given by
\begin{equation}
\label{eq:indexbending}
n = n_{0}\left[1-\kappa(z)y-U(y)\right].
\end{equation}
From Fermat’s principal of least time, the ray equation can by derived~\cite{SalehTeich} for an arbitrary refractive index profile, resulting in
\begin{equation}
\dfrac{d}{ds}\left(n\dfrac{dy}{ds}\right) =\dfrac{\partial n}{\partial y} ~ ~ ~ , ~ ~ ~ ds=\sqrt{dy^2+dz^2}.
\end{equation}
In the approximation that the ray only moves appreciably in the $z$-direction, we may write $ds^2 = dy^2 + dz^2 \approx dz^2.$ Also, it is assumed that we are working in the small bending approximation, thus resulting in $(\partial n/\partial z)(dy/dz)$ being negligible in comparison to $\partial n/\partial y$. This results in the ray equation taking the form
\begin{equation}
\label{eq:rayeq}
\frac{d^2y}{dz^2} = \frac{1}{n}\frac{\partial n}{\partial y}.
\end{equation}
Taking Eq.~(\ref{eq:indexbending}) as our refractive index profile and noting that practical waveguides  satisfy $\kappa(z)y << 1$ and $U(y) << 1$, the ray equation takes the form
\begin{equation}
\label{eq:diff}
\frac{d^2y}{dz^2} = F(y) - \kappa(z),
\end{equation}
where $F(y) = - dU(y)/dy$. We assume that the random variable $\kappa(z)$ has a mean of zero and is white-noise correlated, that is
\begin{eqnarray}
\label{eq:corr}
\big<\kappa(z)\big> &= &0 \, , \nonumber \\
\big<\kappa(z)\kappa(z')\big> & = & \kappa_{0}^2\delta(z-z') \, .
\end{eqnarray}
The above second order stochastic differential equation can be separated into a pair of Langevin equations, given by
\begin{eqnarray}
\label{eq:langevin}
\frac{dy}{dz} & = & \theta(z) \, , \nonumber \\
\frac{d\theta}{dz} & = & F(y) - \kappa(z) \, .
\end{eqnarray}

We may use Eq.~(\ref{eq:LDER}) and Eq.~(\ref{eq:FPDER}) to derive a Fokker-Planck equation which describes the evolution of the probability distribution of the ray along any point of the waveguide. This distribution tells us the probability of the ray being at the position $y$, with angle $\theta$, at the point $z$ along the waveguide. The FPE is given by
\begin{equation}
\left[\dfrac{\partial}{\partial z}
+\theta\dfrac{\partial}{\partial y}
+F\dfrac{\partial}{\partial\theta}
-\dfrac{\kappa_0^2}{2}\dfrac{\partial^2}{\partial\theta^2}\right]P(y,\theta,z)=0.
\label{eq:FokkerPlanck2}
\end{equation}
This FPE looks different from the Schr\"{o}dinger equation, so it is not trivial to see how a SUSY transformation can be applied. However, it is possible to bring this equation to a canonical form, as we will show. First, the FPE depends on two variables $y$ and $\theta$, and we need an equation for one space-like variable (z being the imaginary time here). Noting that in the absence of loss, the Hamiltonian for the system is given by
\begin{align}
E=\dfrac{1}{2}\theta^2+U(y).
\label{eq:Energy}
\end{align}

For our purposes, the absence of loss means that over many ray periods the micro-bending effect should not substantially change the energy of the ray ensemble. We introduce the phase-space action variable~\cite{Goldstein} as the area enclosed in the phase-space of the ray, as the ray traverses one full cycle, and is defined as
\begin{equation}
\label{eq:defIE}
I(E)=\iint d\theta dy=\oint_T\theta dy=T\overline{\theta^2},
\end{equation}
where the period of the ray is denoted $T$ and the average over one full cycle is denoted by the overbar. We will also need the derivative of $I$ with respect to energy, which is given by
\begin{equation}
\label{eq:IprimeE}
I^\prime(E)=\oint_T\dfrac{1}{\sqrt{2(E-U)}}dy=T.
\end{equation}

Now we can use the above action variable and its derivative to write the FPE~(\ref{eq:FokkerPlanck2}) in terms of one space-like variable $E$ and one time-like variable ($z$). The resulting FPE, which is explicitly derived in Appendix~\ref{appendix-a} and its general form is given in Eq.~(\ref{finalfpe}), reads
\begin{equation}
\label{eq:FokkerPlanck3}
\dfrac{2}{\kappa^2_0}\dfrac{\partial P(E,z)}{\partial z}=
-\dfrac{\partial P(E,z)}{\partial E}
+\dfrac{\partial^2[\mathbb{Z}(E)P(E,z)]}{\partial E^2},
\end{equation}
where $P(E,z)$ now represents the probability distribution for the ray to possess energy $E$ when at a distance $z$ along the waveguide, and the variable $\mathbb{Z}$ is defined as 
\begin{equation}
\label{eq:Z}
\mathbb{Z}(E) \equiv \frac{I(E)}{\frac{\partial I(E)}{\partial E}}.
\end{equation}

We now have a FPE which can be related to the Schr\"{o}dinger equation. 

\subsection{Obtaining an index profile from a FPE}

In order to bring Eq.~(\ref{eq:FokkerPlanck3}) into an isomorphic form with respect to the Schr\"{o}dinger equation, we will perform a change of variables on the energy $E$. We now define the energy to be a function of a new variable, denoted $\mathcal{E}$. In order to write the FPE in a way suitable for performing a SUSY transformation, we look for separable solutions of the form $P(E,z)=B(E)Q(E,z)$. Plugging this solution into Eq.~(\ref{eq:FokkerPlanck3}) gives
\begin{eqnarray}
\label{long}
&& \frac{2}{\kappa_{0}^2}B\frac{\partial Q}{\partial z} = -\frac{\partial B}{\partial E}Q - B\frac{\partial Q}{\partial E} + \frac{\partial^2 \mathbb{Z}}{\partial E^2}BQ + \mathbb{Z}\frac{\partial^2 B}{\partial E^2}Q \, \nonumber \\ 
&& \mathbb{Z}B\frac{\partial^2 Q}{\partial E^2} +2(\frac{\partial \mathbb{Z}}{\partial E}\frac{\partial B}{\partial E}Q + \frac{\partial \mathbb{Z}}{\partial E}B\frac{\partial Q}{\partial E}+\mathbb{Z}\frac{\partial B}{\partial E}\frac{\partial Q}{\partial E}) \, .
\end{eqnarray}
Now we choose $E(\mathcal{E})$ and $B(E)$ to satisfy the relationships
\begin{eqnarray}
\label{change}
&& \mathbb{Z}(\frac{{\partial \mathcal{E}}}{\partial E})^{2} = 1 ~ ~ , ~ ~ \mathbb{Z}=(\frac{\partial E}{\partial \mathcal{E}})^2 \, , \\
\label{eq:Breq}
&& 2\mathbb{Z}\frac{\partial B}{\partial E}+\big(\dfrac{3}{2}\frac{\partial \mathbb{Z}}{\partial E}-1\big)B=0 \, .
\end{eqnarray}
The solutions to Eq.~(\ref{change}) and Eq.~(\ref{eq:Breq}) are given as follows
\begin{eqnarray}
\label{eq:EeZ}
&& \mathcal{E}=\int \dfrac{dE}{\sqrt{\mathbb{Z}}} ~ ~ ~ , ~ ~
E=\int\sqrt{\mathbb{Z}}\, d\mathcal{E} \, , \\
\label{eq:Bsol}
&& B=\mathbb{Z}^{-3/4}\exp\Big(\int\dfrac{dE}{2\mathbb{Z}}\Big) \, .
\end{eqnarray}

Note that in Eq.~(\ref{long}), $B$ appears in every term, so an overall constant in Eq.~(\ref{eq:Bsol}) does not effect the physics and thus can be set to a convenient value. Plugging Eq.~(\ref{eq:Bsol}) into Eq.~(\ref{long}) and utilizing the change of variables defined in Eq.~(\ref{change}) to write $\frac{\partial }{\partial E}$ as $\tfrac{1}{\sqrt{\mathbb{Z}}}\frac{\partial }{\partial \mathcal{E}},$
results in the following imaginary-time ($Z=it$) Schr\"{o}dinger equation
\begin{equation}
\label{eq:SEQ}
- \dfrac{\partial Q}{\partial Z}= - \dfrac{\partial^2 Q}{\partial \mathcal{E}^2}
+V(\mathcal{E})Q.
\end{equation}
We have absorbed the coefficient $2/\kappa_{0}^2$ into a redefinition of the partial derivative with respect to $z$, namely $Z=k_0^2z/2$, and the potential energy function is given by 
\begin{equation}
V(\mathcal{E}) = \frac{1}{2}\Big(\frac{\dddot{I}}{\dot{I}}-\frac{1}{2\dot{I}^2}\Big),
\end{equation}
where we have denoted differentiation with respect to $\mathcal{E}$ by $\cdot$. We simplify this expression for the potential energy function by defining a new function $\mathbb{F}$ as follows:
\begin{equation}
\label{eq:ffunction}
\mathbb{F}^4
=\frac{\partial \big[I^2(E)\big]}{\partial E}
=2\Big[\frac{\partial {I}(\mathcal{E})}{\partial \mathcal{E}}\Big]^2.
\end{equation}
The potential now takes the form
\begin{equation}
\label{eq:VFE}
V(\mathcal{E})=\mathbb{F}^{-1} \dfrac{\partial^2}{\partial\mathcal{E}^2} \mathbb{F}.
\end{equation}

We have now arrived at a FPE that can undergo a SUSY transformation. The next step in our goal of finding an index profile with less loss is to perform a SUSY transformation and obtain a partner potential to $V(\mathcal{E})$. Once this is achieved, we use Eq.~(\ref{eq:VFE}) and Eq.~(\ref{eq:ffunction}) to derive the corresponding action variable $I(E)$, and from this we can find the new index profile $U(y)$ by using an Abel transform, which is shown below. From Eq.~(\ref{eq:IprimeE}), we can relate the derivative of the action variable $I$ to the derivative of the index profile through an Abel transform as follows
\begin{eqnarray}
\dfrac{\partial I}{\partial E} & = & \oint_T\dfrac{1}{\sqrt{2(E-U)}}dy=
2\sqrt{2}\int^{U^{-1}(E)}_0\dfrac{1}{\sqrt{E-U}}dy \, \nonumber \\
& = & \int^E_0\left(\dfrac{2\sqrt{2}}{\partial_yU}\right)\dfrac{dU}{\sqrt{E-U}}\, ,
\end{eqnarray}
where in the last step we made a change of variable from $y$ to $U$.
An Abel transform~\cite{Arfken} between two functions $f(x)$ and $g(t)$ is given by
\begin{eqnarray}
f(x) & = & \int_{0}^{x} g(t) \frac{d t}{\sqrt{x-t}} \, , \\
g(t) & = & \frac{1}{\pi} \frac{d}{d t} \int_{0}^{t} f(x) \frac{d x}{\sqrt{t-x}} \, .
\end{eqnarray}
Therefore, we have
\begin{eqnarray}
\label{eq:IprimeE2}
\frac{\partial I(E)}{\partial E} & = & \int^E_0\left(\dfrac{2\sqrt{2}}{\partial_yU}\right)\dfrac{dU}{\sqrt{E-U}} \, , \\
\label{eq:invIprimeE2}
\dfrac{2\pi\sqrt{2}}{\partial_yU} & = & \frac{\partial}{\partial U} \int_{0}^{U} \frac{\partial I(E)}{\partial E} \frac{d E}{\sqrt{U-E}} \, .
\end{eqnarray}
Eq.~(\ref{eq:invIprimeE2}) can be further simplified by taking $U(y=0)=0$ and integrating both sides. This results in
\begin{equation}
\label{eq:invIprimeE22}
y=\dfrac{1}{2\sqrt{2}\pi}\int_{0}^{U} I^\prime(E) \dfrac{d E}{\sqrt{U-E}}.
\end{equation}

Equation~(\ref{eq:invIprimeE22}) allows us to derive a new index profile $U(y)$ from our new action variable. Note that in the first section of the background material, at the beginning of the paper, the SUSY formalism assumes we begin with a potential which has a well behaved ground-state wavefunction in the unbroken-SUSY case, and from this we generate a partner potential with the same spectrum, with the exception of the ground state energy. In our current problem, such a transform would actually make an index profile with more loss. What we must now address is the question regarding whether our starting potential is a $V_+$ potential derivable from a partner potential $V_{-}$, related through a superpotential with unbroken SUSY. We address this question in the latter part of the next section.

\section{Example of a monomial index profile} 

We now apply the above formalism to a specific example. We consider a common~\cite{Gloge,MafiGraded} index profile of the form
\begin{equation}
\label{indexex}
U(y)=\Delta |y/y_c|^\alpha, 
\end{equation}
where $y_c$ is half of the width of the waveguide under consideration. As mentioned previously, the SUSY-QM formalism discussed in Section II starts with a potential $V_-$ and obtains a partner potential $V_+$. Assuming that the superpotential $W$ realizes an unbroken SUSY, this would in fact lead to an index profile with more loss. 
This is always possible by virtue of the superpotential generated from the normalizable ground state associated with the original index profile. On the other hand, finding an index profile that results in less loss assumes that we begin with $V_+$ and obtain a partner potential $V_-$. As we will see, whether this can be achieved depends on the value of the parameter $\alpha$.

\subsection{Obtaining an Index Profile with More Loss}

In order to gain some intuition for the application of the above formalism, we first workout the case of obtaining an index profile which results in more loss. This will give us an example of an index profile which is not as efficient as the monomial profile. We begin here with treating $\alpha$ as a general parameter, but later in this section we will focus on the special case of $\alpha = 2.$
Since the potential in Eq.~(\ref{indexex}) is symmetric under $y\to -y$, we only consider the domain $y\ge 0$ when discussing $U.$ Using Eq.~(\ref{eq:defIE}), we can obtain the action variable $I(E)$ by computing the following integral
\begin{eqnarray}
\label{eq:integrali}
I(E) & = & 2\int_{-y_{c}}^{y_{c}}\sqrt{2[E-\Delta(y/y_c)^\alpha]}dy \, \nonumber \\
& = & \dfrac{\sqrt{8\pi}\, \Gamma(1+1/\alpha)}{\Gamma(3/2+1/\alpha)}
y_c\,\Delta^{-1/\alpha}E^{1/\zeta} \, , 
\end{eqnarray}
where $\zeta = 2\alpha/(\alpha+2)$.

By taking the derivative of $I$ with respect to energy, we see that Eq.~(\ref{eq:Z}) becomes
\begin{equation}
\label{exZ}
\mathbb{Z}(E)=\zeta E.
\end{equation}
Using Eq.~(\ref{eq:EeZ}) we obtain 
\begin{equation}
\label{bcs}
4E=\zeta\mathcal{E}^2,
\end{equation}
and from Eq.~(\ref{eq:Bsol}) we have
\begin{equation}
B = \zeta^{-3/4}E^{1/2\zeta - 3/4} =
\zeta^{-3/4}(\zeta/4)^{1/\zeta}\mathcal{E}^{\frac{1-\alpha}{\alpha}}. 
\end{equation}
As discussed in the paragraph immediately following Eq.~(\ref{eq:Bsol}), $B$ is only defined up to an overall constant. Thus, for simplicity, we choose 
\begin{equation}
\label{actualB}
B = \mathcal{E}^{\frac{1-\alpha}{\alpha}}.
\end{equation}
Next, we obtain the function $\mathbb{F}(\mathcal{E})$ from Eq.~(\ref{eq:ffunction}) as
\begin{equation}
\mathbb{F}(\mathcal{E}) =\pi^{1/4} \left[{\frac{8y_c\Gamma(1+1/\alpha)}{\zeta \Gamma(3/2+1/\alpha)}\Delta^{-1/\alpha}(\zeta / 4)^{1/\zeta}}\right]^{1/2}\mathcal{E}^{1/\zeta -1/2}. 
\end{equation}
We can now use Eq.~(\ref{eq:VFE}) to acquire the potential $V(\mathcal{E})$ as
\begin{equation}
\label{eq:Vsolution}
V(\mathcal{E})
=\dfrac{1-\alpha}{\alpha^2}\dfrac{1}{\mathcal{E}^2}
=\dfrac{\nu^2-1/4}{\mathcal{E}^2},
\end{equation}
where $\nu=(2-\alpha)/2\alpha$.
Note that we have used Eq.~(\ref{eq:integrali}) to write $\zeta$ in terms of the parameter $\alpha.$ 

Now we want to find the ground state wave function of Eq.~(\ref{eq:SEQ}) with the potential given by Eq.~(\ref{eq:Vsolution}) and use this ground state to generate the superpotential, as in Eq.~(\ref{eq:susyground}), and from this obtain the new partner potential. We can find the ground state by an indirect way. We can directly solve Eq.~(\ref{eq:FokkerPlanck3}) in terms of decaying Bessel functions, and then use the substitution $P(E,z) = B(E)Q(E,z)$, which we introduced to derive Eq.~(\ref{long}), along with Eq.~(\ref{eq:Bsol}), to obtain the ground state of Eq.~(\ref{eq:SEQ}). Inserting $\mathbb{Z}$, given in Eq.~(\ref{exZ}), into Eq.~(\ref{eq:FokkerPlanck3}) gives
\begin{equation}
\dfrac{\partial P}{\partial z}=
\Big[\dfrac{\kappa^2_0}{2}(2\zeta-1)
\Big]\dfrac{\partial P}{\partial E}
+\dfrac{\kappa_0^2}{2}\zeta E\dfrac{\partial^2P}{\partial E^2}.
\end{equation}
We now make a change of variable as
\begin{equation}
\label{eq:EepsA}
\mathcal{E}=A\sqrt{E},
\end{equation}
where $\kappa_0^2/2 = 4/\zeta A^2$. This leads to the new form of the FPE in terms of the variable $\mathcal{E}$ as
\begin{equation}
\dfrac{\partial P}{\partial z}=
\dfrac{\partial^2 P}{\partial \mathcal{E}^2}
+
\dfrac{3\zeta-2}{\zeta}
\dfrac{1}{\mathcal{E}}\dfrac{\partial P}{\partial \mathcal{E}}.
\end{equation}
By writing $P=\mathcal{E}^\nu\rho$, and recalling that $\nu = (2 - \alpha)/2 \alpha$, we find
\begin{equation}
\dfrac{\partial \rho}{\partial z}=
\dfrac{\partial^2 \rho}{\partial \mathcal{E}^2}
+\dfrac{1}{\mathcal{E}}\dfrac{\partial\rho}{\partial \mathcal{E}}
-\nu^2\dfrac{1}{\mathcal{E}^2}\rho.
\end{equation}

The stationary solutions to this differential equation can be written as $\rho(z,\mathcal{E})=e^{-\lambda_m z}\rho_m(\mathcal{E})$, which results in
\begin{equation}
\dfrac{\partial^2 \rho_m}{\partial \mathcal{E}^2}
+\dfrac{1}{\mathcal{E}}\dfrac{\partial\rho_m}{\partial \mathcal{E}}
+(\lambda_m-\nu^2\dfrac{1}{\mathcal{E}^2})\rho_m=0.
\end{equation}
The solution is given in terms of Bessel functions as
\begin{equation}
\rho_m=J_\nu(\sqrt{\lambda_m}\mathcal{E})=J_\nu(\sqrt{\lambda_m}A\sqrt{E}).
\end{equation}
From Eq.~(\ref{eq:Energy}), we take the boundary condition to be $\rho_m(E=\Delta)=0$, which is based on the assumption that the loss is small over the waveguide length. This results in
\begin{eqnarray}
\label{nunu}
&& u_{\nu m}=\sqrt{\lambda_m}A\sqrt{\Delta} \, , \nonumber \\
&& \lambda_m=\dfrac{u^2_{\nu m}}{A^2\Delta}=\dfrac{u^2_{\nu m}\zeta \kappa_0^2}{8\Delta}=\dfrac{\alpha}{\alpha+2}u^2_{\nu m}\dfrac{\kappa_0^2}{4\Delta} \, ,
\end{eqnarray}
where $u_{\nu m}$ is the $m$'th root of the $J_\nu$ Bessel function. We therefore obtain the following solution for normal modes of $P$:
\begin{equation}
P_m(\mathcal{E},z)=e^{-\lambda_m z}\mathcal{E}^\nu J_\nu(\sqrt{\lambda_m}\mathcal{E}).
\end{equation}
Now we can use the factorization relation $P(E,z) = B(E)Q(E,z)$. Plugging in Eq.~(\ref{actualB}) gives us the normal mode solutions for $Q$, namely 
\begin{equation}
\label{Qsolution}
Q(\mathcal{E},z) = e^{-\lambda_{m}z}Q_m = e^{-\lambda_{m}z}\mathcal{E}^{1/2}J_\nu(\sqrt{\lambda_m}\mathcal{E}).
\end{equation}
The ground state is given by
\begin{equation}
Q_{0}(\mathcal{E}) = \mathcal{E}^{1/2}J_\nu(\sqrt{\lambda_{0}}\mathcal{E}).
\end{equation}
Calculating the superpotential from Eq.~(\ref{eq:susyground}), we obtain
\begin{equation}
\label{Eq:W-1}
W(\mathcal{E};\nu) = -\dfrac{\nu + 1/2}{\mathcal{E}}+\sqrt{\lambda_{0}}\dfrac{J_{\nu+1}(\sqrt{\lambda_{0}}\mathcal{E})}{J_{\nu}(\sqrt{\lambda_{0}}\mathcal{E})}.
\end{equation}

We must now take into account the fact that the SUSY-QM formalism requires the ground state to be a zero energy eigenstate in order to perform a SUSY transformation. Thus we must actually work with the potential $V_{-} = V(\mathcal{E})-\lambda_0$. This amounts to multiplying the solutions in Eq.~(\ref{Qsolution}) by a factor of $e^{\lambda_{0}z}.$ Physically this amounts to modifying all of the decay constants by the addition of $\lambda_{0}$.

Partner potentials $V_{\pm}$ can be explicitly derived from Eq.~(\ref{eq:vpm}). Inserting Eq.~(\ref{Eq:W-1}) into Eq.~(\ref{eq:vpm}) gives
\begin{equation}
\label{Vm1}
V_{-}(\mathcal{E};
\nu) = \dfrac{\nu^2-1/4}{\mathcal{E}^2}-\lambda_{0}
\end{equation}
and
\begin{eqnarray}
\label{Vp1}
V_{+}(\mathcal{E};\nu) & = & \lambda_0+\dfrac{1}{\mathcal{E}^2}\Big[\dfrac{3}{4}+\nu(\nu+2) + \dfrac{2\lambda_{0}\mathcal{E}^2J^2_{\nu+1}(\sqrt{\lambda_{0}}\mathcal{E})}{J^2_{\nu}(\sqrt{\lambda_{0}}\mathcal{E})} \, \nonumber \\
& - & \dfrac{(4\nu+2)\sqrt{\lambda_0}\mathcal{E}J_{\nu+1}(\sqrt{\lambda_{0}}\mathcal{E})}{J_{\nu}(\sqrt{\lambda_{0}}\mathcal{E})}\Big] \, .
\end{eqnarray}

\begin{figure}[!ht]
\includegraphics[width=0.5\textwidth]{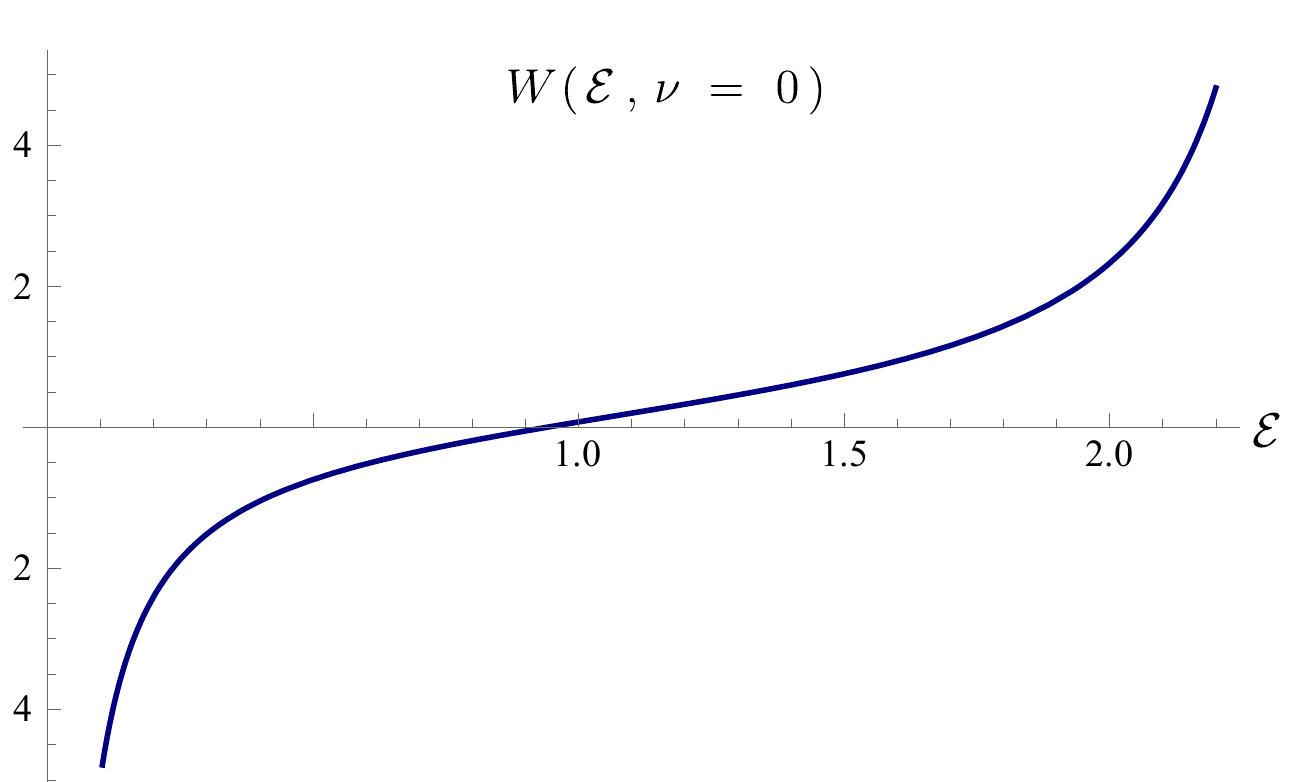}
\caption{Plot of the superpotential $W(\mathcal{E},\nu)$ for the value $\nu=0$.}
\label{fig:susyw}
\end{figure}

\begin{figure}[!ht]
\includegraphics[width=0.5\textwidth]{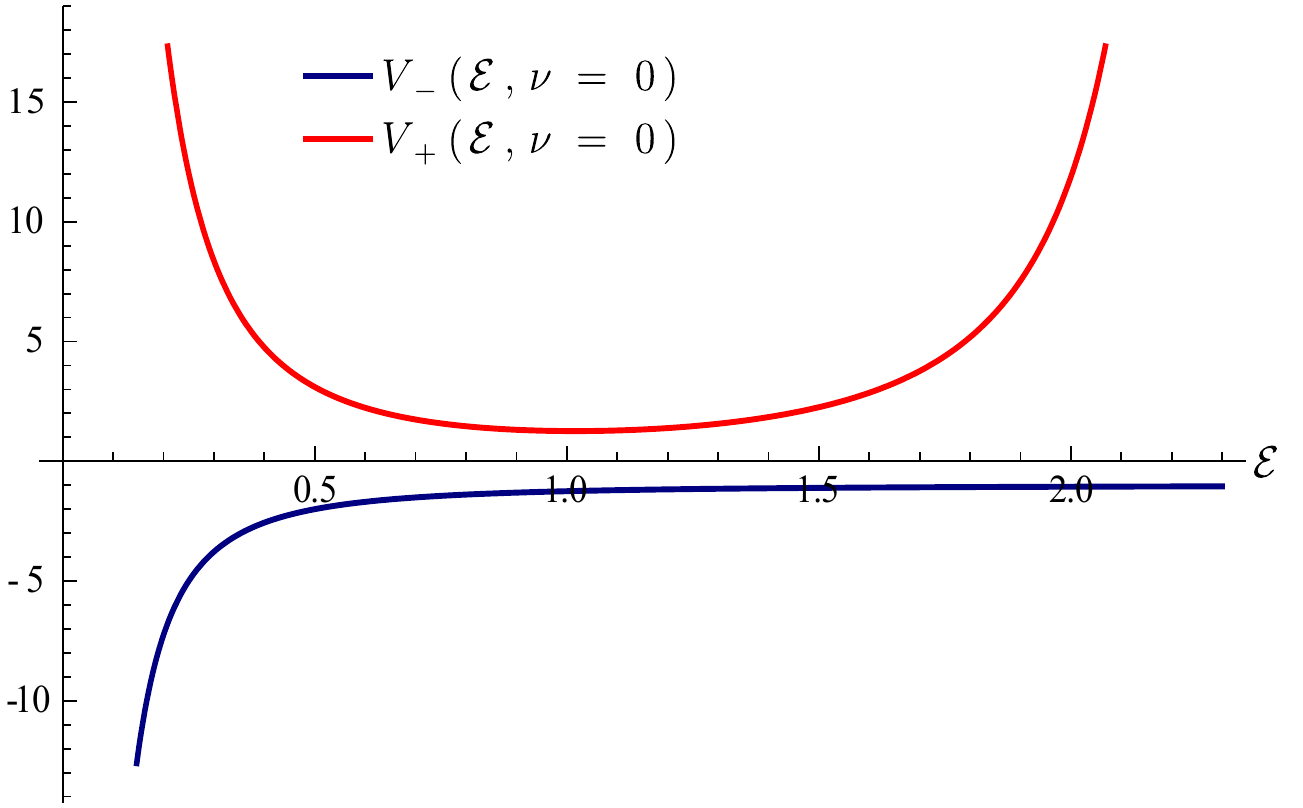}
\caption{Plot of the partner potentials for the value $\nu=0$.}
\label{fig:partners}
\end{figure}

We provide a plot of the superpotential and the partner potentials, up to an overall scaling factor, in Fig.~\ref{fig:susyw} and Fig.~\ref{fig:partners}, respectively, for the value $\nu = 0$. Note that pulling out an overall factor of $\sqrt{\lambda_0}$ in Eq.~(\ref{Eq:W-1}) allows us to use the variable $\sqrt{\lambda_{0}}\mathcal{E}$. Also, since the energy is defined in the domain $0 \leq E \leq \Delta$, the variable $\mathcal{E}$ lies in the domain $0 \leq \mathcal{E} \leq \sqrt{4\Delta /\zeta}$, as can be seen from Eq.~(\ref{bcs}). In passing, we would like to point out that the relationship between the bound states of the two partner potentials in Fig.~\ref{fig:partners} is an example of a system having bound states in the continuum. The idea of bound states being found in the continuum was first introduced by John von Neumann and Eugene Wigner~\cite{vw}. Since then the idea of such states have been applied to many physical systems~\cite{bsc}.

As discussed previously in Section II, we can see from the figure that the superpotential for $\nu=0$ does in fact realize an unbroken SUSY. From here on we take a definite value for the parameter $\alpha$ in order to further our example. We take $\nu=0$ and from Eq.~(\ref{eq:EepsA}) we get $\alpha=2$. We can now proceed in determining the index profile $U(y)$ for the partner potential $V_{+}(\mathcal{E};\nu=0).$

In order to obtain an index profile we must solve Eq.~(\ref{eq:VFE}) with the above $V_+$  partner potential, given in Eq.~(\ref{Vp1}), with $\nu = 0.$ Thus we must solve
\begin{eqnarray}
\label{eq:f}
\mathbb{F}^{-1} \dfrac{\partial^2}{\partial\mathcal{E}^2} \mathbb{F} & = &   
\lambda_{0} +\frac{1}{\mathcal{E}^2}\Big[\frac{3}{4} - \frac{2\sqrt{\lambda_{0}}\mathcal{E}J_{1}(\sqrt{\lambda_{0}}\mathcal{E})}{J_{0}(\sqrt{\lambda_{0}\mathcal{E}})} \, \nonumber \\
& + & \frac{2\lambda_{0}\mathcal{E}^2J_{1}(\sqrt{\lambda_{0}}\mathcal{E})^2}{J_{0}(\sqrt{\lambda_{0}\mathcal{E}})^2}\Big] \, .
\end{eqnarray}
Clearly, this differential equation should be solved by numerical methods. In order to get a numerical solution, we must specify the boundary conditions which $\mathbb{F}$ will satisfy. If we focus on solving Eq.~(\ref{eq:f}) near the origin we can approximate $V_+$ as $V_{+} \approx 1/\mathcal{E}^2.$ From this approximation we may obtain a solution for $\mathbb{F}$ near the origin by solving
\begin{equation}
\frac{d \mathbb{F}^2}{d\mathcal{E}^2} = \frac{\mathbb{F}}{\mathcal{E}^2}.
\end{equation}
The solution of this equation is of the form $c_{1}\mathcal{E}^{3/2} + c_{2}\mathcal{E}^{-1/2}.$ We set the coefficient $c_2$ to zero. This is reasonable since physically $\mathbb{F}^4$ represents the kinetic energy of the ray over one cycle and should go to zero as the energy goes to zero, as can be seen from Eq.~(\ref{eq:defIE}) and Eq.~(\ref{eq:ffunction}). So, near the origin, we find $\mathbb{F}(\mathcal{E})\propto \mathcal{E}^{3/2}.$ From this solution we see that
$\dot{\mathbb{F}}(\mathcal{E})\propto (3/2)\mathcal{E}^{1/2}$. We now have the necessary tools to set the two boundary conditions required to solve Eq.~(\ref{eq:f}). From the solution $\mathbb{F}$ and its derivative near the origin, we require $\mathbb{F}(0) = \dot{\mathbb{F}}(0) = 0.$

Now that we are able to obtain a solution for Eq.~(\ref{eq:f}), we can use Eq.~(\ref{eq:ffunction}) to obtain the derivative of the action variable as 
\begin{equation}
\label{eq:in}
\partial I(\mathcal{E})/\partial \mathcal{E} = \mathbb{F}^{2}/\sqrt{2}. 
\end{equation}
We require the action variable to obey $I(0) = 0.$ From this boundary condition we can integrate Eq.~(\ref{eq:in}) to get an expression for $I(\mathcal{E})$. We can now obtain an expression for $\mathbb{Z}(\mathcal{E)}$ from Eq.~(\ref{eq:Z}) using $\sqrt{Z}\partial/\partial E = \partial/\partial \mathcal{E}$, we see 
\begin{equation}
\label{eq:zfin}
\mathbb{Z}(\mathcal{E})=2I^2/\mathbb{F}^4.
\end{equation}
Inserting $Z(\mathcal{E})$ into Eq.~(\ref{eq:EeZ}) gives us the relationship between $\mathcal{E}$ and $E$, allowing us to write an expression for $I(E)$ from $I(\mathcal{E})$. Now that we have $I(E)$ and its derivative $\partial I(E)/\partial E$, we may use Eq.~(\ref{eq:invIprimeE22}) to derive the new index profile $U(y)$.

\begin{figure}[!ht]
\includegraphics[width=0.48\textwidth]{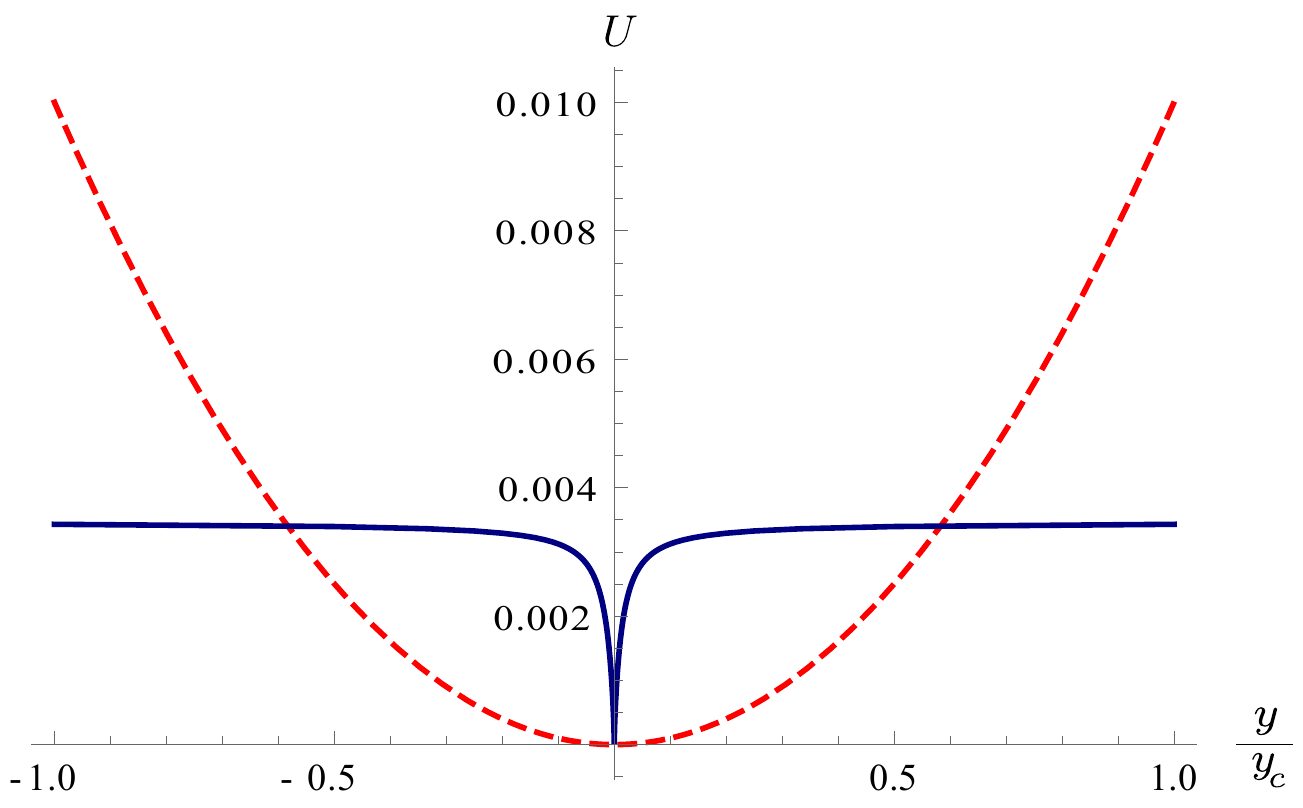}
\caption{The dashed line plot shows the original index profile $U=\Delta(y/y_c)^2$ (corresponding to $V_-$), while the solid line plot shows the new index profile $U$ (corresponding to $V_+$).}
\label{fig:potentials}
\end{figure}

From Eq.~(\ref{eq:Z}) we see that $\mathbb{Z}$ is only defined up an overall scaling factor, due to it being a ratio of a function and its derivative. Thus, when we use Eq.~(\ref{eq:zfin}) and Eq.~(\ref{eq:invIprimeE22}) to generate the new index profile, we only get the profile up to an overall scaling factor. We will choose the scaling factor such that the $y$ dependence of the index profile is normalized  to $y_{c}$, as was done with the index profile (\ref{indexex}) which we started with. We plot the original index profile and new index profile, generated numerically by following the above procedure, in Fig.~\ref{fig:potentials}.

\subsection{Lowering the Loss}
Now that we have seen an example of using the above formalism to derive a new index profile, we move on to the more practical issue of finding a profile which has lower loss than the one given in Eq.~(\ref{indexex}). As noted in the beginning of this section, in order to obtain an index profile which exhibits less loss, we must assume that we are starting with a potential $V_{+}$ which is derivable from a partner potential $V_{-}$ through an unbroken SUSY superpotential. We now derive the condition that must be satisfied for this identification of the partner potentials to be true. 

Assuming that $V(x) = V_{+}$ means that there exists a superpotential $W$ such that $V_{+} = W^2 + W'$, where we have set $\hbar/\sqrt{2m} = 1$. This superpotential is generated from the ground state, $\psi_{0}$, of the partner potential $V_{-}$ via $W = -\psi_{0}'/\psi_{0}$. Plugging this expression into our formula for $V_{+}$ gives
\begin{equation}
\label{eq:condition}
\frac{d^2\psi_{0}}{dx^2} + \psi_{0}V_{+} = \frac{2}{\psi_{0}}(\frac{d\psi_{0}}{dx})^2.
\end{equation}

This tells us that for an unbroken SUSY partner potential $V_{-}$ to exist, the above equation for $\psi_{0}$ must have a normalizable solution. In the absence of such a normalizable $\psi_{0}$, we will not be able to write down a partner potential $V_-$. Lastly, we digress to point out that SUSY-QM can also be used to generate a family of potentials that are completely isospectral to a given potential $V(x)$, this results in a class of index profiles that present equal loss. In the SUSY-QM literature this method goes by the name of Isospectral Deformation\cite{Gango}. We provide a brief example of this in Appendix~\ref{appendix-b}.

We now have the necessary tools needed to determine whether an index profile with less loss can be derived from the monomial profile. We see from Eq.~(\ref{eq:condition}) and Eq.~(\ref{eq:Vsolution}) that for this to be the case there must exist a normalizable solution to the equation
\begin{equation}
\frac{d^2\psi_{0}}{dx^2} + \frac{\nu^{2}-1/4}{x^2}\psi_{0} = \frac{2}{\psi_{0}}(\frac{d\psi_{0}}{dx})^2.
\end{equation}
The solution to this equation is of the form
\begin{equation}
\psi_{0}(x) = c_{2}\exp\Big[\frac{1}{2}\Big((\sqrt{4\nu^2} - 1)\log{(x)} - 2\log(x^{2\nu}+c_1)\Big)\Big].
\end{equation}

In order for $\psi_{0}(x)$ to be normalizable, we see from the first argument in the exponential that $\nu$ must satisfy $|\nu| > 1/2.$ Therefore, this tells us that given the index profile in Eq.~(\ref{indexex}), the possible range of $\alpha$ which permits a less-lossy partner profile is given by $\vert \frac{2-\alpha}{2\alpha} \vert > 1/2,$ and thus forcing $\alpha < 1.$ We now can see that the choice of $\alpha = 2$ for the index profile in Eq.~(\ref{indexex}) cannot lead to an index profile with less loss, using the SUSY-QM formalism. For $\alpha < 1$, the previous formalism can be applied to derive the corresponding refractive index profile.

\section{Conclusion} 

We address the problem of mitigating micro-bending attenuation in a waveguide, by way of a supersymmetric quantum mechanical approach. Starting from an initial refractive index profile, and taking the micro-bending into account, it is shown how to derive an associated Fokker-Planck equation, which describes the probability of finding a light ray with a particular energy at some distance along the waveguide. The Fokker-Planck equation is then related to the Schr\"{o}dinger equation, and the methods of SUSY-QM are applied, resulting in a new Fokker-Planck equation. The procedure for generating a new refractive index profile from the corresponding Fokker-Planck equation, is then developed. It is shown that it is always possible to derive a refractive index which results in more loss, but not always possible to derive a refractive index resulting in less loss. We show that the existence of a SUSY transformation, resulting in a partner index profile with less loss, is determined by the solutions of a non-linear differential equation. We apply our formalism to the monomial index profile, with exponent $\alpha$ treated as a free parameter, and show that a refractive index profile with less loss may be derived only when $\alpha < 1.$

It should be noted that while this work focuses on a problem in classical stochastic optics, the analysis can be applied to other stochastic problems that are based on the use of the Fokker-Planck equation. Furthermore, the developed procedure in this work could be expanded to deal with an optical fiber, where the micro-bending would no longer be confined to one dimension. Interesting applications may be found by applying the above formalism to integrated optical waveguide structures.
   
\section{Acknowledgement}

This research is supported by grant number W911NF-19-1-0352 from the United States Army Research Office.

\appendix
\section{From Langevin to Fokker-Planck}
\label{appendix-a}

The following equation is the ray equation describing the trajectory of a light ray through a waveguide when subjected to micro-bending and a velocity dependent damping term. In our problem we neglected the damping term, but here we include it for completeness.
\begin{equation} \label{eq:1}
\frac{d^2y}{dz^2} = -\gamma \frac{dy}{dz}-\frac{dU(y)}{dy}-\kappa(z)
\end{equation}

Assuming low energy loss over many cycles, the energy is approximately given by
\begin{equation} \label{eq:2}
E = \frac{\dot{y}^2}{2}+U(y).
\end{equation}
Differentiating this equation with respect to $z$ gives
\begin{equation}
\dot{E} = \dot{y}\ddot{y}+\frac{dU(y)}{dy}\dot{y} = (\ddot{y}+\frac{dU(y)}{dy})\dot{y}.
\end{equation}
Using (\ref{eq:1}) we have 
\begin{equation}
\dot{E} =-(\gamma\dot{y}+\kappa(z))\dot{y}.
\end{equation}
So our two Langevin equations are given by
\begin{eqnarray}
&& \dot{y} = \theta = \pm \sqrt{2(E-U(y))} \, , \nonumber\\
&& \dot{E} = -\gamma \theta^2 - \kappa(z) \theta \, .
\end{eqnarray}

Now we want to use these Langevin equations to write a FPE. In general, when given a set of Langevin equations of the form
\begin{equation}
\frac{\partial y_i}{\partial t} = A_i(\textbf{y},t) + \sum_{k}B_{ik}(\textbf{y},t)\zeta_k(t),
\end{equation}
with mean and auto-correlation given as
\begin{eqnarray}
\big<\zeta_k(t)\big> &= &0 \, , \nonumber \\
\big <\zeta_{i}(t_{1})\zeta_{j}(t_{2}) \big > & = & 2D\delta_{ij}\delta{(t_{1}-t_{2})}\, ,
\end{eqnarray}

the corresponding FPE is of the form
\begin{eqnarray}
&& \frac{\partial P(\textbf{y},t)}{\partial t} = \, \nonumber \\
&& - \sum_i\frac{\partial}{\partial y_i}\{[A_i(\textbf{y},t)+ D\sum_{jk}B_{jk}(\textbf{y},t)\frac{\partial B_{ik}(\textbf{y},t)}{\partial y_i}]P\} \, \nonumber \\
&& + D\sum_{ij}\frac{\partial^2}{\partial y_i \partial y_j}\{[\sum_kB_{ik}(\textbf{y},t)B_{jk}(\textbf{y},t)]P\} \, .
\end{eqnarray}
For our particular Langevin equations, we get the following FPE:
\begin{eqnarray}
&& \frac{\partial P(y,E,z)}{\partial z} = -\frac{\partial}{\partial y} (\theta P(y,E,z)) + \gamma \frac{\partial}{\partial E} (\theta^2 P(y,E,z)) \, \nonumber \\
&& - \frac{\kappa_{0}^2}{2} \frac{\partial P(y,E,z)}{\partial E} + \frac{\kappa_{0}^2}{2}\frac{\partial^2}{\partial E^2}[\theta^2 P(y,E,z)] \, .
\end{eqnarray}
We now have a FPE which gives the probability density $P=P(z,E,y)$. We want to reduce this equation to one that depends only on $z$ and $E$.

Let us separate the probability density into two parts, i.e., $P(y,E,z) = P(E,z)P(y|E,z).$ The probability to find the particle in the interval $y$ and $y + dy$ simply is $dt/T$, where $dy = \vert \theta \vert$ and $T$ is the period of oscillations for energy $E$. Note that we have used $\vert \theta \vert$ as the probability density, which is positive definite. The conditional probability density is then given by
\begin{equation}
P(y|E,z) = {1 \over \vert \theta \vert T}.    
\end{equation}
It is easy to verify that $P(y|E,z)$ is normalized.

With the help of Eq.~(A10), we can now rewrite the FPE in Eq.~(A9) in terms of $P(E,z)$ as
\begin{eqnarray}
&&{\partial \over \partial z} \left({P(E,z) \over \theta T}\right) = - {\partial \over \partial y} \left({P(E,z) \over T}\right) + \gamma \frac{\partial}{\partial E}\left({\theta P(E,z) \over T}\right)  \, \nonumber \\
&& - \frac{\kappa_{0}^2}{2} {\partial \over \partial E} \left({P(E,z) \over \theta T} \right) + \frac{\kappa_{0}^2}{2}\frac{\partial^2}{\partial E^2}\left({\theta P(E,z)\over T}\right) \, .
\end{eqnarray}
The first term on the right-hand side of this equation vanishes. The second term is
\begin{equation}
\gamma \frac{\partial}{\partial E}\left({\theta P(E,z) \over T}\right) = \gamma \theta {\partial \over \partial E} \left({P(E,z) \over T}\right) + \gamma {P(E,z) \over \theta T} .      
\end{equation}
The last term can be expanded as
\begin{eqnarray}
&& \frac{\kappa_{0}^2}{2}\frac{\partial^2}{\partial E^2}\left({\theta P(E,z)\over T}\right) = {\kappa_0^2 \over 2} {\partial \over \partial E}\left({P(E,z) \over \theta T}\right) \, \nonumber \\
&& + {\kappa_0^2 \over 2} {1 \over \theta} {\partial \over \partial E} \left({P(E,z) \over T}\right) + {\kappa_0^2 \over 2} \theta {\partial^2 \over \partial^2 E} \left({P(E,z) \over T}\right) \, . \nonumber \\
\end{eqnarray}
The left-hand side of Eq.~(A11) can be written as
\begin{equation}
{\partial \over \partial z} \left({P(E,z) \over \theta T}\right) = {-{\dot \theta} \over \theta^2 T} P(E,z) + {1 \over \theta T} {\partial P(E,z) \over \partial z}.     
\end{equation}

After multiplying both sides of Eq.~(A14) with $\theta T$, and using Eq.~(A12) and Eq.~(A13), we find
\begin{eqnarray}
&& {- {\dot \theta} \over \theta} P(E,z) + {\partial P(E,z) \over \partial z} \, \nonumber \\
&& = \gamma P(E,z) + \gamma \theta^2 T {\partial \over \partial E} \left({P(E,z) \over T}\right)+ \ \nonumber \\
&& {\kappa_0^2 \over 2} T {\partial \over \partial E} \left({P(E,z) \over T}\right) + {\kappa_0^2 \over 2} \theta^2 T {\partial^2 \over \partial^2 E} \left({P(E,z) \over T}\right).
\end{eqnarray}
We now average both sides of this equation over one period of oscillation with energy $E$ and denote differentiation with respect to $E$ by $\prime.$ Noting that the average of ${\dot \theta} / \theta$ vanishes, we have
\begin{eqnarray}
&& {\partial P(E,z) \over \partial z} = \gamma P(E,z) + \gamma I {\partial \over \partial E} \left({P(E,z) \over I^{\prime}}\right) \, \nonumber \\
&& {\kappa_0^2 \over 2} I^{\prime} {\partial \over \partial E}\left({P(E,z) \over I^{\prime}}\right) + {\kappa_0^2 \over 2} I {\partial^2 \over \partial^2 E } \left({P(E,z) \over I^{\prime}}\right) \, , \nonumber \\
&& \,
\end{eqnarray}
where we have used
\begin{equation}
I = {\overline {\theta^2}}T,    
\end{equation}    
and
\begin{equation}
I^{\prime} = T.
\end{equation}
Note that
\begin{equation}
\gamma P(E,z) + \gamma I {\partial \over \partial E} \left({P(E,z) \over I^{\prime}}\right) = \gamma {\partial \over \partial E} \left({{I \over I^{\prime}} P(E,z)} \right) ,        
\end{equation}
and
\begin{eqnarray}
&& {\kappa_0^2 \over 2} I^{\prime} {\partial \over \partial E}\left({P(E,z) \over I^{\prime}}\right) + {\kappa_0^2 \over 2} I {\partial^2 \over \partial^2 E } \left({P(E,z) \over I^{\prime}}\right) \, \nonumber \\
&& = {\kappa_0^2 \over 2} {\partial^2 \over \partial E^2} \left[{I \over I^{\prime}} P(E,z) \right] - {\kappa_0^2 \over 2} {\partial \over \partial E} P(E,z).
\end{eqnarray}
We therefore arrive at the following FPE for $P(E,z)$
\begin{eqnarray}
\label{finalfpe}
{\partial P(E,z) \over \partial z} & = & {\partial \over \partial E} \left[ \left(\gamma {I \over I^{\prime}} - {\kappa_0^2 \over 2} \right) P(E,z) \right] \, \nonumber \\ 
& + & {\kappa_0^2 \over 2} {\partial^2 \over \partial^2 E} \left[ {I \over I^{\prime}} P(E,z) \right] \, . 
\end{eqnarray}



\newpage
\begin{thebibliography}{10}
\newcommand{\enquote}[1]{``#1''}

\bibitem{Witten1}
E.~Witten, \enquote{Dynamical breaking of supersymmetry,}
Nucl. Phys. B \textbf{188}, 513--554 (1981).

\bibitem{Witten2}
E.~Witten, \enquote{Supersymmetry and Morse theory,} J. Diff. Geom. \textbf{17}, 661--692 (1982).

\bibitem{Cooper1}
F.~Cooper, A.~Khare and U.~Sukhatme, \enquote{Supersymmetry and Quantum Mechanics,} Phys. Rept. \textbf{251}, 267--385 (1995).


\bibitem{Cooper2}
F.~Cooper and B.~Freedman, \enquote{Aspects of supersymmetric quantum
  mechanics,} Ann. Phys. \textbf{146}, 262--288 (1983).

\bibitem{Cooper3}
F.~Cooper, A.~Khare, and U.~Sukhatme, \emph{Supersymmetry in quantum mechanics}
  (World Scientific, Singapore, 2001).
  
\bibitem{Suk85b}
C.~V.~Sukumar, \enquote{Supersymmetric quantum mechanics of one-dimensional systems,}
J. Phys. A \textbf{18}, L57--L61 (1985).  
  
\bibitem{Gango}  
A.~Gangopadhyaya, J.~V.~Mallow, C.~Rasinariu, \enquote{Supersymmetric Quantum Mechanics An Introduction} (World Scientific, Singpore, 2010)

\bibitem{Jafarizadeh}
M.~A.~Jafarizadeh and H.~Fakhri, \enquote{Supersymmetry and shape invariance in differential equations of mathematical physics,} Phys. Lett. A \textbf{230}, 164--170 (1997).

\bibitem{Dutt}
R.~Dutt, A.~Khare, and U.~P.~Sukhatme, \enquote{Supersymmetry, shape invariance, and exactly solvable potentials,} Am. J. Phys. \textbf{56}, 163--168 (1988).

\bibitem{Dutt2}
R.~Dutt, A.~Khare, and U.~P.~Sukhatme, \enquote{Supersymmetry-inspired WKB approximation in quantum mechanics,} Am. J. Phys. \textbf{59}, 723--727 (1991).

\bibitem{Dutt3}
R.~Adhikari, R.~Dutt, A.~Khare, and U.~P.~Sukhatme, \enquote{Higher-order WKB approximations in supersymmetric quantum mechanics,} Phys. Rev. A \textbf{38}, 1679 (1988).

\bibitem{scatter}
T.~Shimbori, T.~Kobayashi, \enquote{Supersymmetric Quantum Mechanics of Scattering,} Physics Letters B. \textbf{501}, 245-248. 

\bibitem{Sukumar2}
C.~V.~Sukumar, \enquote{Supersymmetric quantum mechanics and the inverse scattering method,} J. Phys. A \textbf{18}, 2937 (1985).

\bibitem{Chumakov}
S.~M.~Chumakov and K.~B.~Wolf, \enquote{Supersymmetry in {H}elmholtz optics,}
  Phys. Lett. A \textbf{193}, 51--53 (1994).
  
\bibitem{Ganainy2012}
R.~El-Ganainy, K.~G.~Makris, and D.~N.~Christodoulides,
\enquote{Local PT invariance and supersymmetric parametric oscillators,}
Phys. Rev. A \textbf{86}, 033813 (2012)

\bibitem{Miri2013}
M.~A.~Miri, M.~Heinrich, R.~El-Ganainy, and D.~N.~Christodoulides, \enquote{Supersymmetric optical structures,} Phys. Rev. Lett. \textbf{110}, 233902 (2013).

\bibitem{Miri2013-2}
M.~A.~Miri, M.~Heinrich, and D.~N.~Christodoulides, \enquote{Supersymmetry-generated complex optical potentials with real spectra,}
Phys. Rev. A \textbf{87}, 043819 (2013).

\bibitem{Heinrich2014}
M.~Heinrich, M.~A.~Miri, S.~St\"{u}tzer, R.~El-Ganainy, S.~Nolte, A.~Szameit, and D.~N.~Christodoulides, \enquote{Supersymmetric mode converters,} Nat. Commun. \textbf{5}, 3698 (2014).

\bibitem{Heinrich2014-2}
M.~Heinrich, M.~A.~Miri, S.~St\"{u}tzer, R.~El-Ganainy, S.~Nolte, A.~Szameit, and D.~N.~Christodoulides, \enquote{Supersymmetric mode converters,} Nat. Commun. \textbf{5}, 3698 (2014).  

\bibitem{Heinrich2014-3}
M.~Heinrich, M.~A.~Miri, S.~St\"{u}tzer, S.~Nolte, D.~N.~Christodoulides, and A.~Szameit, \enquote{Observation of supersymmetric scattering in photonic lattices,} Opt. Lett. \textbf{39}, 6130--6133 (2014).

\bibitem{Miri2014}
M.~A.~Miri, M.~Heinrich, and D.~N.~Christodoulides, \enquote{SUSY-inspired one-dimensional transformation optics,} Optica \textbf{1}, 89--95 (2014).

\bibitem{Ganainy2015}
R.~El-Ganainy, L.~Ge, M.~Khajavikhan, and D.~N.~Christodoulides, \enquote{Supersymmetric laser arrays,} Phys. Rev. A \textbf{92}, 033818 (2015).

\bibitem{Walasik2018}
W.~Walasik, B.~Midya, L.~Feng, and N.~M.~Litchinitser, \enquote{Supersymmetry-guided method for mode selection and optimization in coupled systems,} Opt. Lett. \textbf{43}, 3758--3761 (2018).

\bibitem{Midya2018}
B.~Midya, W.~Walasik, N.~M.~Litchinitser, and L.~Feng, \enquote{Supercharge optical arrays,} Opt. Lett. \textbf{43}, 4927--4930 (2018).

\bibitem{Midya2019}
B.~Midya, H.~Zhao, X.~Qiao, P.~Miao, W.~Walasik, Z.~Zhang, N.~M.~Litchinitser, and L.~Feng, \enquote{Supersymmetric microring laser arrays,} Photon. Res. \textbf{7}, 363--367 (2019).

\bibitem{Walasik2019}
W.~Walasik, N.~Chandra, B.~Midya, L.~Feng, and N.~M.~Litchinitser, \enquote{Mode-sorter design using continuous supersymmetric transformation,} Opt. Express \textbf{27}, 22429--22438 (2019).

\bibitem{Zhong2019}
Q.~Zhong, S.~Nelson, M.~Khajavikhan, D.~N.~Christodoulides, and R.~El-Ganainy, \enquote{Bosonic discrete supersymmetry for quasi-two-dimensional optical arrays,} Photon. Res. \textbf{7}, 1240--1243 (2019).

\bibitem{Hokmabadi2019}
M.~P.~Hokmabadi, N.~S.~Nye, R.~El-Ganainy, D.~N.~Christodoulides, and M.~Khajavikhan, \enquote{Supersymmetric laser arrays,} Science \textbf{363}(6427), 623--626 (2019).
  
\bibitem{Junker}
G.~Junker, \emph{Supersymmetric methods in quantum and statistical physics,}
(Springer-Verlag, Berlin, 2012).

\bibitem{Morales}
J.~Morales, and J.~J.~Peña, \enquote{Supersymmetric quantum mechanics and statistical physics: Schr\"{o}dinger-like thermodynamic equation,} Phys. Scr. \textbf{74}, 71 (2006).

\bibitem{Polotto}
F.~Polotto, E.~D.~Filho, and J.~Chahine, R.~J.~ de Oliveira, \enquote{Supersymmetric quantum mechanics method for the Fokker–Planck equation with applications to protein folding dynamics,} Physica \textbf{493}, 286--300 (2018). 

\bibitem{PaSo82}
G.~Parisi, and N.~Sourlas, \enquote{Supersymmetric field theories and stochastic differential equations,} Nucl. Phys. B \textbf{206}, 321--332 (1982).

\bibitem{Zirnbauer}
M.~R.~Zirnbauer, \enquote{The supersymmetry method of random matrix theory,} arXiv preprint math-ph/0404057 (2004).

\bibitem{Efetov}
K.~Efetov, \emph{Supersymmetry in disorder and chaos,} (Cambridge University Press, New York, 1999).

\bibitem{PaSo79}
G.~Parisi, and N.~Sourlas, \enquote{Random magnetic fields, supersymmetry, and negative dimensions,} Phys. Rev. Lett. \textbf{43}, 744 (1979).

\bibitem{Gardner}
W.~B.~Gardner, \enquote{Microbending loss in optical fibers,} Bell Labs Tech. J. \textbf{54}, 457--465 (1975).

\bibitem{Jay}
J.~A.~Jay, \enquote{An overview of macrobending and microbending of optical fibers,} White paper of Corning,  1--21 (2010).

\bibitem{Paper}
S.~Ward, R.~Allahverdi, and A.~Mafi, \enquote{Supersymmetric analysis of stochastic micro-bending in optical waveguides,} arXiv:2009.11847v2; Accepted for publication at OSAC (2021).

\bibitem{Rousseau}  
M.~Rousseau and J.~Arnaud, \enquote{Ray theory of microbending,} Opt. Commun. \textbf{25}, 333--336 (1978).  

\bibitem{Arnaud}  
J.~Arnaud and M.~Rousseau, \enquote{Ray theory of randomly bent multimode optical fibers,} Opt. Lett. \textbf{3}, 63--65 (1978).

\bibitem{ArnaudII}
M.~Rousseau, and J.~Arnaud, \enquote{Microbending loss of multimode square-law fibres: a ray theory,} Electron. Lett. \textbf{9}(13), 265--267 (1977).

\bibitem{ArnaudIII}
F.~de Fornel, J.~Arnaud, and P.~Facq, \enquote{Microbending effects on monomode light propagation in multimode fibers,} J. Opt. Soc. Am. \textbf{73}, 661-668 (1983).

\bibitem{Han}
Z.~Han, P.~Zhang, and S.~I.~Bozhevolnyi, "Calculation of bending losses for highly confined modes of optical waveguides with transformation optics," Opt. Lett. \textbf{38}, 1778-1780 (2013).

\bibitem{ChrisI}
M.~A.~Miri, M.~Heinrich, R.~El-Ganainy, D.~N.~Christodoulides, \enquote{Supersymmetric optical structures,} Phys. Rev. Lett. \textbf{110}, (23) 233902 (2013).

\bibitem{ChrisII}
M.~Heinrich, M.~A.~Miri, S.~Stützer, et al, \enquote{Supersymmetric mode converters.} Nat. Commun. \textbf{5}, 3698 (2014).

\bibitem{ChrisIII}
M.~A.~Miri, M.~Heinrich, R.~El-Ganainy, D.~N.~Christodoulides, \enquote{Supersymmetry-generated complex optical potentials with real spectra,}
Phys. Rev. A \textbf{87}, 043819 (2013).

\bibitem{HeinrichI}
M.~Heinrich, M.~A.~Miri, S.~Stutzer, S.~Nolte, D.~N.~Christodoulides, and A.~Szameit, \enquote{Observation of supersymmetric scattering in photonic lattices,} Optics Letters, \textbf{39}(21), 6130--6133 (2014).

\bibitem{Sergey}
S.~M..~Chumakov and K.~BernardoWolf, \enquote{Supersymmetry in Helmholtz optics,} Mod. Phys. Lett., \textbf{193}(1), 51--53 (1994).

\bibitem{Stefano}
S.~Longhi, \enquote{Supersymmetric transparent optical intersections,} Opt. Lett. \textbf{40}, 463--466 (2015)

\bibitem{Laba}
H.~P.~Laba and V.~M.~Tkachuk, \enquote{Quantum-mechanical analogy and supersymmetry of electromagnetic wave modes in planar waveguides,}
Phys. Rev. A \textbf{89}, 033826 (2014).

\bibitem{Walasik}
W.~Walasik, B.~Midya, L.~Feng, and N.~M.~Litchinitser, \enquote{Supersymmetry-guided method for mode selection and optimization in coupled systems,} Opt. Lett. \textbf{43}, 3758-3761 (2018)

\bibitem{Rosu}
H.~C.~Rosu, \enquote{Supersymmetric Fokker-Planck strict isospectrality,} Phys. Rev. E \textbf{56}, 2269 (1997).

\bibitem{PolottoII}
F.~Polotto et al, \enquote{Solutions of the Fokker–Planck equation for a Morse isospectral potential,}  J. Phys. A: Math. Theor. \textbf{43}, 015207 (2009).

\bibitem{Bern}
M.~Bernstein and L.~S.~Brown, \enquote{Supersymmetry and the Bistable Fokker-Planck Equation,} Phys. Rev. Lett. \textbf{52}, 1933 (1984).

\bibitem{FP}
H.~Risken and T.~Frank, \enquote{The Fokker-Planck Equation,} (Springer-Verlag, Berlin Heidelberg, 1996).

\bibitem{MafiAcoustic}
R.~A.~Herrera, C.~T.~Law, and A.~Mafi, \enquote{Calculation of the
acousto-optic coupling coefficients in optical fibers,} Opt. Commun.
\textbf{305}, 217--220 (2013).

\bibitem{Heiblum}
M.~Heiblum and J.~Harris, \enquote{Analysis of curved optical waveguides by
conformal transformation,} IEEE J. Quantum Electron. \textbf{11}, 75--83 (1975).

\bibitem{Menachem}
Z. Menachem, \enquote{Propagation in curved waveguides and applications,} J. Electromagn. Waves Appl. \textbf{33}(14), 1801--1833 (2019).

\bibitem{Sheehan}
R.~N.~Sheehan, S.~Horne and F.~H.~Peters, \enquote{The design of low-loss curved waveguides,} Opt Quantum Electron. \textbf{40}, 1211-–1218 (2008).

\bibitem{SalehTeich}
B.~E.~A.~Saleh and M.~C.~Teich, \emph{Fundamentals of photonics,} (John Wiley \& sons, New Jersey, 2019).

\bibitem{Goldstein}
H.~ Goldstein, \emph{Classical Mechanics,} (Addison-Wesley, Cambridge ,1980).

\bibitem{Arfken}
G.~B.~Arfken, H.~J.~Weber and F.~E.~Harris, \emph{Mathematical Methods for Physicists A Comprehensive Guide,} (Academic Press, New York, 2012).

\bibitem{Gloge}
D.~Gloge and E.~A.~J.~Marcatili, \enquote{Multimode theory of graded-core fibers,} Bell. Syst. Tech. J. \textbf{52}, 1563--1578 (1973).

\bibitem{MafiGraded}
A.~Mafi, \enquote{Pulse propagation in a short nonlinear graded-index multimode optical fiber,} J. Lightw. Technol. \textbf{30}, 2803--2811 (2012).
  
\bibitem{vw}
J. von Neumann and E. Wigner, Zeit. Phys. \textbf{30}, (1929) 465.

\bibitem{bsc}
C.W.~Hsu, B,~Zhen, A.~Douglas Stone, et al., \enquote{Bound states in the continuum,} Nat. Rev. Mater. \textbf{1}, 16048 (2016).

\end{thebibliography}
\section{Isospectral Deformation}
\label{appendix-b}

Given a partner potential $V_{-}(\mathcal{E})$, there exists a class of potentials with the same spectrum and the same partner $V_{+}(x)$, given by
\begin{equation}
\label{iso}
\hat{V}_{-}(\lambda,\mathcal{E}) = V_{-}(\mathcal{E})-2\frac{d}{d\mathcal{E}}\ln{[I(\mathcal{E})+\lambda]},
\end{equation}
where 
\begin{equation}
\label{iso2}
I(\mathcal{E}) = \int_{-\infty}^{\mathcal{E}} dt \psi_{0}^2(t),
\end{equation}
and $\lambda$ is an integration constant that is allowed to be in the range $\lambda < -1$ or $\lambda > 0.$ As an example, we look at the potential generated from the monomial index profile with exponent given by $\alpha = 2$,
\begin{equation}
\label{original}
V_{-} = -\lambda_{0}+\dfrac{\nu^2-1/4}{\mathcal{E}^2},
\end{equation}
which has a ground state given by 
\begin{equation}
\psi_{0} = \sqrt{\mathcal{E}}J_{\nu}(\sqrt{\lambda_{0}}\mathcal{E}).
\textbf{}
\end{equation}

\begin{figure}[H]
\includegraphics[width=0.5\textwidth]{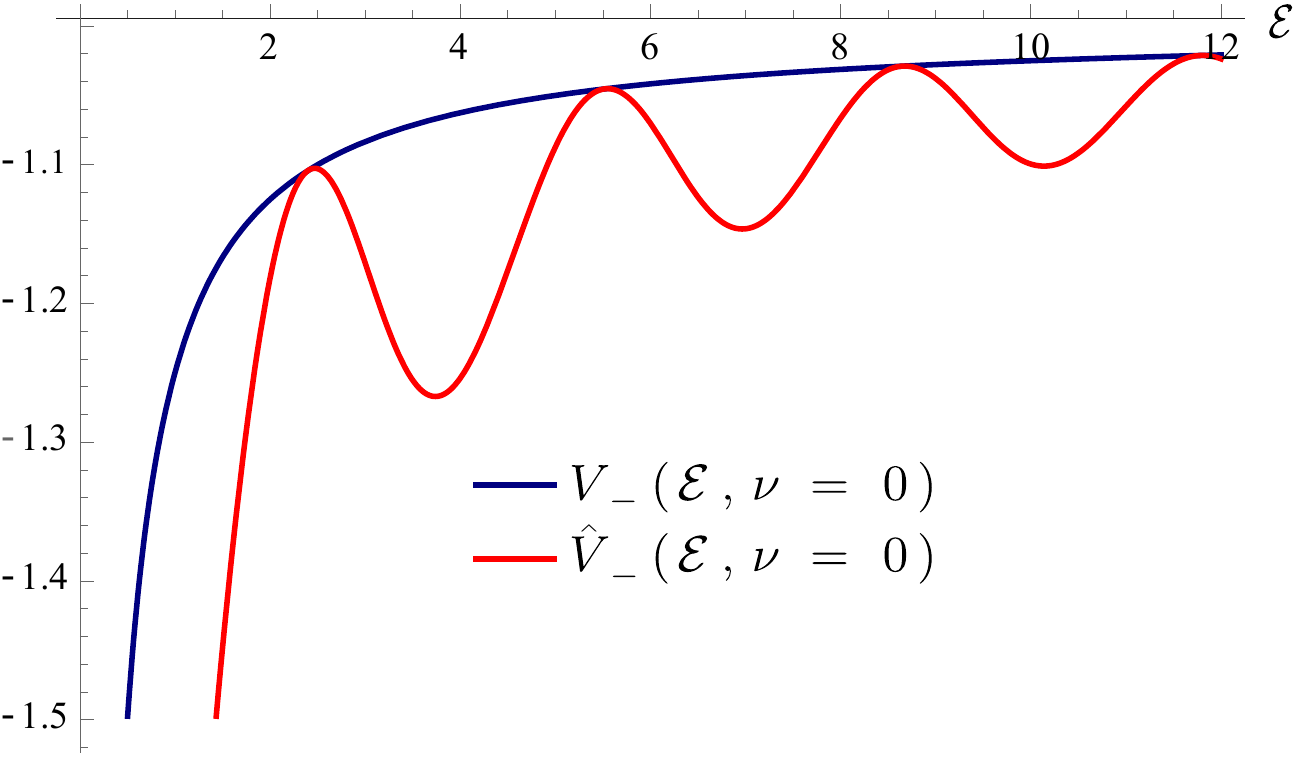}
\caption{Plot of the original potential $V_-$ and the isospectral potential $\hat{V}_{-}$ with parameters $\lambda_0 = 1,\lambda = 1.$}
\label{fig:isospectral}
\end{figure}

Now we derive an isospectral partner potential $\hat{V}_{-}$, by using Eq.~(\ref{iso}) and Eq.~(\ref{iso2}), as follows:
\begin{eqnarray}
&& I = \frac{\mathcal{E}^2}{2\lambda_{0}}[J_{\nu}^2 -J_{\nu -1}J_{\nu+1}] \, , \\
&& \frac{d}{d\mathcal{E}}\ln{[I(\mathcal{E})+\lambda]} =
\frac{A\mathcal{E}^2 + 2B\mathcal{E}}{\mathcal{E}^2[J_{\nu}^2-J_{\nu-1}J_{\nu+1}]+\lambda} \, ,
\end{eqnarray}
where
\begin{eqnarray}
A & = & \sqrt{\lambda_{0}}(J_{\nu}(J_{\nu-1}-J_{\nu+1})-0.5J_{\nu+1}(J_{\nu-2}-J_{\nu}) \, \nonumber \\
& - &0.5J_{\nu-1}(J_{\nu}-J_{\nu+1})),
\end{eqnarray}
and
\begin{equation}
B = J_{\nu}^2 - J_{\nu-1}J_{\nu+1}.
\end{equation}
This results in
\begin{equation}
\hat{V}_{-}(\lambda,\mathcal{E}) = -\lambda_{0}+\dfrac{\nu^2-1/4}{\mathcal{E}^2}\\
- 2\frac{A\mathcal{E}^2 + 2B\mathcal{E}}{\mathcal{E}^2[J_{\nu}^2-J_{\nu-1}J_{\nu+1}]+\lambda}.
\end{equation}
For the special case of $\nu = 0$, we have 
%
%
%
\begin{equation}
I = \frac{\mathcal{E}^2}{2\lambda_{0}}[J_{0}^2 +J_{1}^2],
\end{equation}
and
\begin{equation}
\frac{d}{d\mathcal{E}}\ln{[I(\mathcal{E})+\lambda]} =
\frac{J_{0}^2}{\mathcal{E}(J_{0}^2+J_{1}^2)+\lambda},
\end{equation}
which results in
\begin{equation}
\label{iso3}
\hat{V}_{-}(\lambda,\mathcal{E}) =  -\lambda_{0}-\dfrac{1/4}{\mathcal{E}^2} - \frac{2J_{0}^2}{\mathcal{E}(J_{0}^2+J_{1}^2)+\lambda}.  
\end{equation}
Choosing the parameter values $\lambda_{0} = 1$ and $\lambda = 1$, we plot the isospectral potential (Eq.~(\ref{iso3})) and the original potential (Eq.~(\ref{original})) in Fig~\ref{fig:isospectral}.

Performing an isospectral deformation on a potential $V,$ which is associated with an index profile $U,$ and using the methods discussed previously in this paper will result in the generation of a family of index profiles which have the same loss as the original profile. Further details on the method of isospectral deformation can be found in~\cite{Gango}.

\end{document}